\documentclass[aps,prb,twocolumn,floatfix,showpacs,superscriptaddress,fleqn]{revtex4-1}
\usepackage{amssymb}

\usepackage{graphicx}
\usepackage{amsmath}
\usepackage{bm}
\usepackage{color}
\usepackage{siunitx}

\usepackage[firstpage]{draftwatermark}
\usepackage[breaklinks]{hyperref}
\usepackage[all]{hypcap}
\newcommand*{\neper}{\mathop{}\!\mathrm{e}}
\newcommand*{\boltzmannconstant}{\mathop{}\!\mathrm{k_{\mathrm{B}}}}
\newcommand*{\diff}{\mathop{}\!\text{d}}
\hypersetup{
    plainpages=false,
    bookmarks=false,         
    unicode=false,          
    pdfborder={0 0 0}
    pdftoolbar=true,        
    pdfmenubar=true,        
    pdffitwindow=false,     
    pdfstartview={FitH},    
    pdftitle={BPrevised},    
    pdfauthor={Paulina Plochocka},     
    pdfproducer={LNCMI-CNRS-UJF-UPS-INSA}, 
    pdfkeywords={High} {Magnetic} {Fields}, 
    pdfnewwindow=true,      
    linktoc=section,
    colorlinks=true,       
    linkcolor=blue,          
    citecolor=red,        
    filecolor=magenta,      
    urlcolor=blue           
}

\begin{document}

\title{Excitons in atomically thin black phosphorus}

\author{A.\ Surrente}
\affiliation{Laboratoire National des Champs Magn\'etiques Intenses,
UPR 3228, CNRS-UGA-UPS-INSA, Grenoble and Toulouse, France}

\author{A.\ A.\ Mitioglu}\altaffiliation[Present address: ]{High Field Magnet Laboratory (HFML-EMFL), Institute for Molecules and Materials, Radboud University, Toernooiveld 7, 6525 ED Nijmegen, The Netherlands}
\affiliation{Laboratoire National des Champs Magn\'etiques Intenses,
UPR 3228, CNRS-UGA-UPS-INSA, Grenoble and Toulouse,
France}\affiliation{Institute of Applied Physics, Academiei Str.\ 5,
Chisinau, MD-2028, Republic of Moldova}

\author{K.\ Galkowski}
\affiliation{Laboratoire National des Champs Magn\'etiques Intenses,
UPR 3228, CNRS-UGA-UPS-INSA, Grenoble and Toulouse, France}

\author{W.\ Tabis}
\affiliation{Laboratoire National des Champs Magn\'etiques Intenses,
UPR 3228, CNRS-UGA-UPS-INSA, Grenoble and Toulouse, France}
\affiliation{AGH University of Science and Technology, Faculty of Physics and Applied Computer Science, Al.\ Mickiewicza 30, 30-059 Krakow, Poland}

\author{D.\ K.\ Maude}
\affiliation{Laboratoire National des Champs Magn\'etiques Intenses,
UPR 3228, CNRS-UGA-UPS-INSA, Grenoble and Toulouse, France}

\author{P.\ Plochocka}\email{paulina.plochocka@lncmi.cnrs.fr}
\affiliation{Laboratoire National des Champs Magn\'etiques Intenses,
UPR 3228, CNRS-UGA-UPS-INSA, Grenoble and Toulouse, France}


\date{\today}

\begin{abstract}
Raman scattering and photoluminescence spectroscopy are used to investigate the optical properties of single layer
black phosphorus obtained by mechanical exfoliation of bulk crystals under an argon atmosphere. The Raman spectroscopy,
performed in situ on the same flake as the photoluminescence measurements, demonstrates the single layer character of
the investigated samples. The emission spectra, dominated by excitonic effects, display the expected in plane
anisotropy. The emission energy depends on the type of substrate on which the flake is placed due to the different
dielectric screening. Finally, the blue shift of the emission with increasing temperature is well described using a two
oscillator model for the temperature dependence of the band gap.
\end{abstract}

\maketitle


Black phosphorus, the most stable of all the allotropes of phosphorus, has been intensively studied by different
experimental methods from the early fifties of the last
century.\cite{Keyes53,warschauer1963electrical,Maruyama81,Asahina83,Jamieson63,Wittig68,Sugai85} Bulk black phosphorus
is a semiconductor, with a band gap of about \SI{0.335}{\eV}.\cite{Keyes53,Asahina83} The orthorombic bulk crystal has
a layered structure, with atomic layers bound by weak van der Waals interactions. A single atomic layer is puckered,
with the phosphorus atoms being parallel in the (010) plane.\cite{Narita83,Takao81,Maruyama81} Atomically thin
monolayers have been recently isolated using mechanical exfoliation,\cite{Castellanos14} adding black phosphorus to the
rapidly growing family of emerging two dimensional materials. The band gap of black phosphorus is always direct and can
be tuned from \SI{0.3}{\eV} to the nearly visible part of the spectrum\cite{Tran14,ling2015}. In contrast, graphene is
gapless,\cite{Novoselov05} and the transition metals dichalcogenides (TMDs) have an indirect gap in bulk phase and only
monolayer TMDs have a direct gap.\cite{Wang12} Moreover, black phosphorus exhibits a strong in-plane
anisotropy\cite{Tran14,Low14,xia14,Qiao14,ling2015}, absent in graphene and TMDs. Additionally, the relatively high
mobilities measured at room temperature combined with the direct band gap result in an on/off ratio for FET transistors
of the order of 10$^{5}$. \cite{ling2015,li14,Liu14}

Although black phosphorus has a wide range of possible applications, including tunable photodetectors\cite{Buscema15},
field effect transistors\cite{xia14,ling2015,li14,Liu14} or photon polarizers\cite{Tran14,Low14,xia14,Qiao14,ling2015},
many of its electronic properties are not yet fully understood. The main difficulty arises from the sensitivity of
black phosphorus to its environment, notably its high reactivity when exposed to air and laser
light.\cite{Castellanos14,Koenig14,Gan15} Theoretical calculations predict a band gap of monolayer black
phosphorus ranging from \SI{1}{\eV} \cite{Liu14,Tran14} to \SI{2.15}{\eV} \cite{Castellanos14,Tran14}, with a binding
energy of the neutral exciton of \SI{0.8}{\eV} in vacuum \cite{Tran14} and of \SI{0.38}{\eV} when placed on a SiO$_{2}$
substrate \cite{Castellanos14}. The measured photoluminescence (PL) emission energy from bilayers black phosphorus is
between \SI{1.2}{\eV} \cite{Zhang14} and \SI{1.6}{\eV} \cite{Castellanos14}, while monolayer black phosphorus shows
neutral exciton emission around \SI{1.3}{\eV} \cite{wang2015highly}, \SI{1.45}{\eV} \cite{Liu14} or \SI{1.76}{\eV}
\cite{Yang15}. The latter value was related to the simultaneous observation of charged exciton emission at around
\SI{1.62}{\eV}.\cite{Yang15} Using scanning tunneling microscopy the band gap of a single layer of black phosphorus was
estimated to be \SI{2.05}{\eV}.\cite{liang2014electronic} The exciton binding energy and consequently the emission
energy strongly depend on the dielectric environment (substrate) \cite{Castellanos14,Woomer15}. This could partially
explain the wide range of values for the black phosphorus emission energy found in the literature, possibly related
also to a slightly different composition of the SiO$_2$ substrates employed.
Moreover, owing to the limited life time of the samples, the various
characterization techniques used to identify monolayer black phosphorus (\emph{e.g.\ }atomic force microscopy, Raman
spectroscopy) could not always be performed \emph{on the same flake} where the optical response was investigated.


In this paper we present a systematic investigation of the optical properties of monolayers of black phosphorus. We
analyze the properties of the emission as a function of the dielectric constant of the substrate, excitation power,
polarization and temperature. The single layer character of the investigated flakes is demonstrated using \emph{in
situ} Raman measurements on the same flake used for the PL. We show that the PL emission energy of monolayer black
phosphorus depends on the substrate used. The PL spectra, dominated by excitonic effects, exhibit the expected in plane
anisotropy. Finally, the blue shift of the emission energy with an increasing temperature is well described with a
two-oscillator model for the temperature dependence of the band gap.

Single and few layer black phosphorus flakes have been obtained by mechanical exfoliation of a bulk crystal, purchased
from Smart Elements (99.998\% nominal purity). The mechanical exfoliation was performed in a glove box filled with
argon (Ar) gas ($<\SI{1}{ppm}$ O$_2$, $<\SI{1}{ppm}$ H$_2$O). The flakes were subsequently transferred onto a Si
substrate, in most cases capped with a \SI{300}{\nano\m} thick layer of SiO$_2$ (Si/SiO$_2$ hereafter). The samples
were stored in vials, in the glove box, before their transfer to the cryostat under an Ar atmosphere.

For the optical measurements, the samples were mounted on the cold finger of a He-flow cryostat, which was then rapidly
pumped to a pressure below $\SI{1e-4}{\milli\bar}$, minimizing the exposure of the black phosphorus to air. The
excitation was provided by a frequency-doubled diode laser, emitting at \SI{532}{\nano\m} and focused on the sample by
a 50$\times$ microscope objective (0.55 numerical aperture), yielding a spot size of $\sim\SI{1}{\micro\m}$. The PL and
Raman signals were collected through the same objective and analyzed by a spectrometer equipped with a liquid nitrogen
cooled Si CCD camera.

Raman spectroscopy, which has been shown to be a very precise tool for the determination of the number of layers in
TMDs \cite{Lee10,Gutierrez13,Li12a,Li12,Wang12} and in graphene, \cite{Ferrari06,Gupta06} can also be used to identify
monolayer black phosphorus. Bulk black phosphorus belongs to the $D_{2h}^{18}$ space group. Of the 12 normal modes at
the $\Gamma$ point of the Brillouin zone, six are Raman active.\cite{Sugai81,Sugai85} The two B$_{3g}$ modes are
forbidden in back scattering configuration and the B$_{1g}$ mode at \SI{194}{\centi\m^{-1}} is very weak. Therefore,
only three Raman modes are expected for bulk black phosphorus: A$_{g}^{1}$, A$_{g}^{2}$ and B$_{2g}$ at around
\SI{365}{\centi\m^{-1}}, \SI{470}{\centi\m^{-1}} and \SI{442}{\centi\m^{-1}}, respectively.\cite{Sugai81,Sugai85} The
A$_{g}^{2}$ and B$_{2g}$ modes are related to the in plane vibrations of the atoms, while A$_{g}^{1}$ mode is related
to the out of plane movement of the atoms~\cite{Sugai81,Sugai85}.

\begin{figure}
\begin{center}
\includegraphics[width=1.0\linewidth]{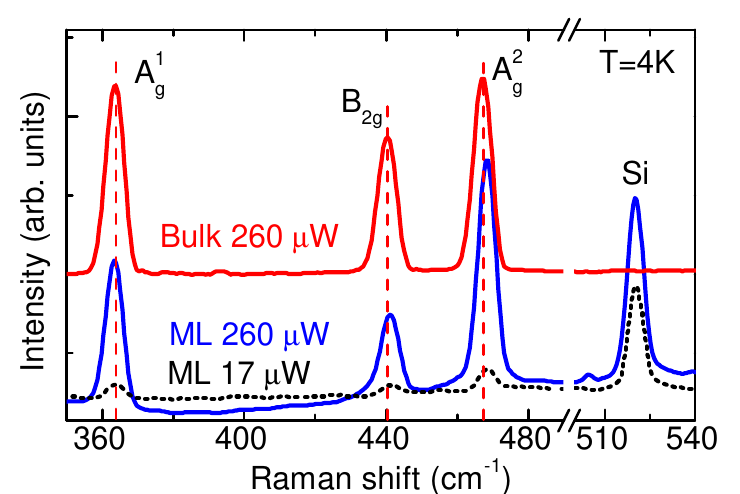}
\end{center}
\caption{(Color online) Typical \si{\micro}Raman spectra of bulk and single layer black phosphorus on Si/SiO$_2$ measured at $T=\SI{4}{\K}$
for different excitation powers as indicated. The dashed lines indicate the position of the three Raman modes
A$_{g}^{1}$, A$_{g}^{2}$ and B$_{2g}$ in bulk black phosphorus.}\label{Figure_Raman}
\end{figure}

Typical calibrated micro Raman (\si{\micro}Raman) spectra of the bulk and single layer black phosphorus are presented
in Fig.\ \ref{Figure_Raman}. For the bulk crystal three strong peaks are observed at \SI{363.7}{\centi\m^{-1}},
\SI{440.3}{\centi\m^{-1}} and \SI{467.5}{\centi\m^{-1}} corresponding well to the expected main Raman modes. A further
peak was systematically observed at \SI{520}{\centi\m^{-1}} (Raman mode of the Si substrate), confirming the correct
calibration of our Raman setup. The slight shift towards lower frequencies, compared to the previously cited literature
values, can be related to the low temperature (\SI{4}{\K}) at which our measurements have been carried
out.\cite{Zhang14} Under low excitation power (\SI{17}{\micro\W}), initially used to avoid any risk of inducing damage
by exposure to the laser light, the Raman spectrum of the monolayer flake is similar to that of the bulk crystal. This
suggests that the monolayer black phosphorus remains crystalline and no oxidization occurred during the transfer of the
sample from the glove box to the cryostat. Compared to bulk, the A$_{g}^{1}$ mode does not shift within experimental
error. Although this Raman mode was found to soften in a monolayer sample \cite{ Liu14}, it was generally found to be
rather insensitive to the number of layers.\cite{Favron15, lu2014plasma} In contrast, the position of A$_{g}^{2}$ and
B$_{2g}$ Raman modes shift to higher frequencies in the monolayer flake. The A$_{g}^{2}$ mode shifts by
$1.7\pm$\SI{0.2}{\centi\m^{-1}} while the B$_{2g}$ shifts by 1.0$\pm$\SI{0.2}{\centi\m^{-1}}. The change of the energy
of the Raman modes as a function of the number of layers in black phosphorus has been already shown for both high
\cite{Favron15,Castellanos14,Liu14} and low frequency modes \cite{Luo15}. The shifts we measure are in good agreement
with the reported values for single layer black phosphorus ($\sim\SI{1.9}{\centi\m^{-1}}$ and around
$\sim\SI{1.0}{\centi\m^{-1}}$ for A$_{g}^{2}$ \cite{Favron15,Castellanos14,Liu14,lu2014plasma} and B$_{2g}$
\cite{Favron15,Castellanos14}, respectively) demonstrating the single layer character of the investigated flake. Raman
measured at high excitation power (\SI{260}{\micro\W}, the same as used for the bulk crystal) does not show any
additional features; the peak positions remain the same as for the data obtained for low excitation power. Thus there
is no sign of laser induced chemical modification of our samples,\cite{Castellanos14} which is important since the
excitation intensity used here is comparable to that used in our micro PL (\si{\micro}PL) spectroscopy measurements.

\begin{figure}[htb]
    \begin{center}
        \includegraphics[width=1.0\linewidth]{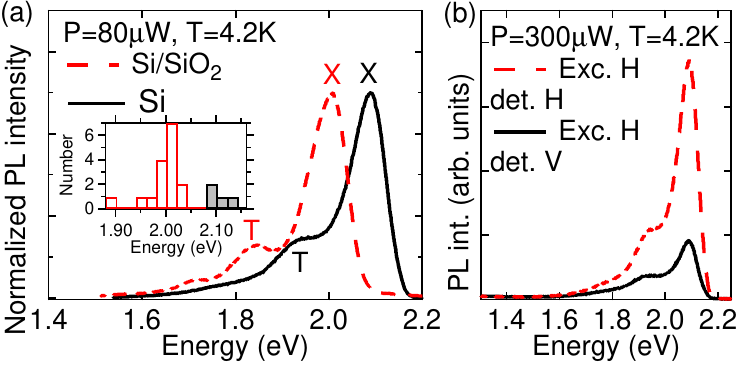}
    \end{center}
\caption{(Color online) (a) Low temperature \si{\micro}PL spectra of black phosphorus flakes on different substrates. X
labels the neutral exciton transition, T the charged exciton. The insert shows a histogram of the emission energy for
the 21 flakes investigated. The mean emission energy is \SI{1.991}{\eV} with a standard deviation $\sigma=\SI{0.036}{\eV}$ for flakes
on a Si/SiO$_2$ substrate and \SI{2.102}{\eV} with $\sigma=\SI{0.015}{\eV}$ for flakes on a Si substrate. (b) Polarization resolved
\si{\micro}PL spectra of a black phosphorus flake on Si.} \label{fig:PL_substrate}
\end{figure}

In Fig.~\ref{fig:PL_substrate}(a) we show representative low temperature \si{\micro}PL spectra for the monolayer flake
previously characterized using Raman spectroscopy. The measurement was performed without removing the flake from the
cryostat, which was maintained under vacuum. The spectrum is dominated by a very strong emission line centered at
around \SI{2}{\eV} identified with neutral exciton recombination. This line is consistently observed at this energy for
all flakes transferred to Si/SiO$_2$ substrates. The typical full width at half maximum is of
$\lesssim\SI{90}{\milli\eV}$ (line broadening induced by scattering with vacancies and impurities in the exfoliated
flakes). The observed value of the exciton recombination energy is relatively high as compared to the theoretical band
gap of $2.15$\,eV computed for black phosphorus single layer,\cite{Castellanos14,Tran14} suggesting that the
investigated sample is indeed a single layer. This is somewhat larger than the values already reported for black
phosphorus single layers of $\eqsim 1.3-\SI{1.76}{\eV}$ \cite{Liu14,Yang15,wang2015highly}. This could be partly
ascribed to the different stoichiometry of the SiO$_2$ layers used as a substrate, which induces a shift in the
emission because of the different dielectric constant of the surrounding medium.\cite{choi2007solvatochromism} To
verify this hypothesis, we have transferred black phosphorus flakes onto a Si substrate covered by a thin
($\sim\SI{2}{\nano\m}$) layer of native oxide. Such a substrate has a larger dielectric constant than the standard
Si/SiO$_2$ substrates. Thus, the exciton binding energy is expected to be lower and the emission energy should be blue
shifted. A typical \si{\micro}PL spectrum of a flake transferred onto a Si substrate is shown in
Fig.~\ref{fig:PL_substrate}(a). While the main spectral features resemble those of a flake on a Si/SiO$_2$ substrate,
there is a large shift of the emission energy of $\sim\SI{80}{\milli\eV}$. This confirms the dependence of the emission
energy on the dielectric environment, in agreement with previous results.\cite{Castellanos14,Woomer15} The effect of
the dielectric environment seems to be stronger than for atomically thin TMDs, in agreement with theory and
experiment.\cite{Lin14} However, we note that the effect of the dielectric environment alone is too small to account
for the large variation of the emission energy reported in the literature.\cite{Liu14,Yang15,wang2015highly}

In the spectrum measured on a Si/SiO$_2$ substrate, in addition to the neutral exciton recombination, we observed
additional features at $\sim\SI{1.84}{\eV}$, which can be attributed to charged exciton recombination (systematically
present in all the investigated flakes)\cite{Yang15} together with a low energy peak, typically at
$\sim\SI{1.72}{\eV}$, and possibly related to excitons bound to impurities. The low energy feature is not always
present, see \emph{e.g.\ }the \si{\micro}PL spectrum on a Si substrate in Fig.~\ref{fig:PL_substrate}(a).


Because of the low symmetry of the crystal structure and the screening in black phosphorus \cite{Tran14,
li2014electrons}, the exciton wave function is expected to be squeezed along the armchair direction, resulting in a
polarization dependent PL emission \cite{wang2015highly}. The linear polarization dependence of the emission of our
flakes was investigated by mounting polarizers/analyzers in both the excitation and the detection paths. For any fixed
orientation of the analyzer in the detection path, the detected signal is always stronger for horizontally (H) polarized excitation. This stems from the polarization dependent
absorption properties of black phosphorus \cite{xia14}. Regardless of the polarization of the excitation beam, whenever
the analyzer in the detection path was set to H, the intensity of the detected signal was a maximum, which
is consistent with the strongly anisotropic nature of the exciton in black phosphorus \cite{wang2015highly}. For
example, in Fig.~\ref{fig:PL_substrate}(b), we show \si{\micro}PL spectra excited with a H polarized
laser light and detected with either H or vertical (V) direction of the analyzer.

\begin{figure}[htb]
    \begin{center}
        \includegraphics[width=1.0\linewidth]{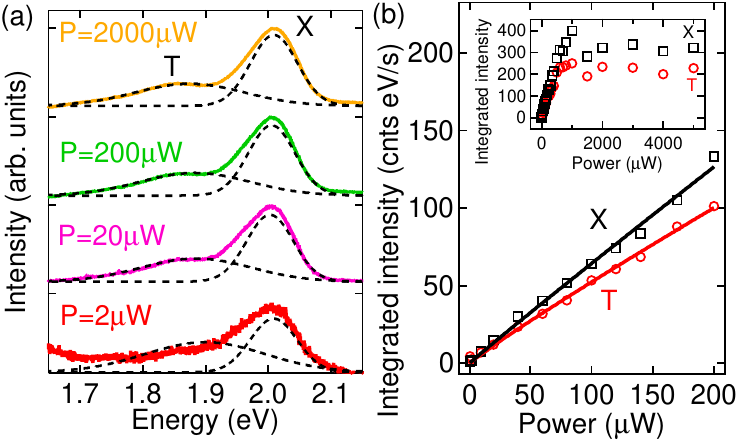}
    \end{center}
\caption{(Color online) (a) \si{\micro}PL spectra of a black phosphorus single layer on Si/SiO$_2$ measured at
different excitation powers. The broken curves are Gaussian fits used to calculate the integrated intensity. (b)
Integrated intensity of the charged (T) and neutral (X) exciton emission versus excitation power. The solid lines are fits to a power law described in the text.} \label{fig:Pdependence}
\end{figure}

The evolution of the \si{\micro}PL spectra as a function of the excitation power gives information concerning the nature of
the observed transition lines. We have measured the power dependence of the \si{\micro}PL spectra for several black
phosphorus monolayer flakes. In Fig.~\ref{fig:Pdependence}(a) we show representative spectra measured at low,
intermediate, and high excitation powers $P$ at $T=\SI{4.2}{\K}$. Even after exciting with $P>\SI{1}{\milli\W}$, the
emission efficiency did not decrease for any of the investigated samples, suggesting that our preparation method helps
improve the stability of the exfoliated black phosphorus flakes. Two peaks appear in the spectra of
Fig.~\ref{fig:Pdependence}(a) for all excitation powers used. The neutral exciton peak slightly blue shifts ($\lesssim\SI{2}{\milli\eV}$) at high excitation powers, possibly due to a localized heating effect induced by the high power
of the incoming laser beam (see also below for a discussion of temperature dependence of the black phosphorus PL).

In Fig.~\ref{fig:Pdependence}(b) we present the dependence of the integrated PL intensity $I$ of neutral and charged
excitons as a function of the excitation power. At low excitation power, both intensities increase linearly with the
power (indicating the absence of biexciton emission), as demonstrated by the fits to the power law $I\propto P^n$,
with $n=0.97\pm0.03$ for the neutral exciton and $n=0.93\pm0.03$ for the charged exciton. At higher excitation power, both excitonic lines saturate at approximately the same level of excitation power [$P=\SI{1}{\milli\W}$, see insert to Fig.\ \ref{fig:Pdependence}(b)],
confirming that the two observed transitions are related to the recombination of a single electron-hole pair.

\begin{figure}[h!]
    \begin{center}
        \includegraphics[width=1.0\linewidth]{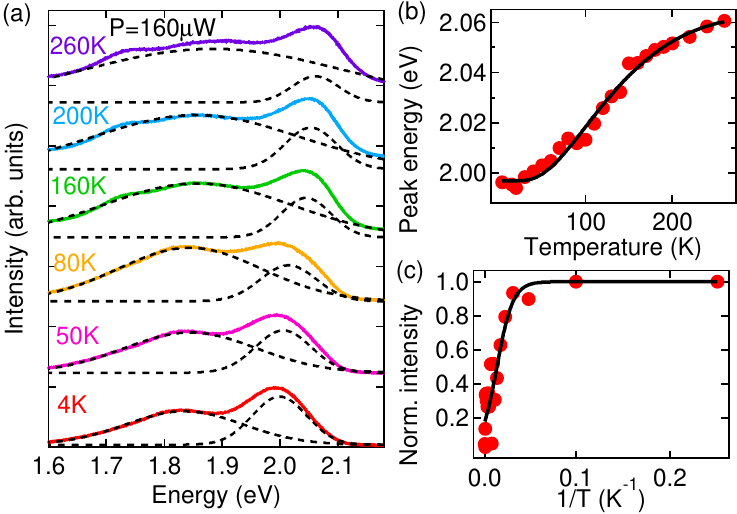}
    \end{center}
\caption{(Color online) (a) \si{\micro}PL spectra of a black phosphorus single layer on Si/SiO$_2$ measured at
different temperatures. The broken lines are Gaussian fits used to extract the emission energy. (b) Emission energy of
the neutral exciton as a function of the temperature. (c) Integrated intensity of the neutral exciton versus
temperature. The solid lines are fits to the models described in the text.} \label{fig:Tdependence}
\end{figure}

In view of the potential use of black phosphorus in a wide variety of electronic and optoelectronic applications, we have investigated the temperature dependence of its optical properties. In Fig.\ \ref{fig:Tdependence}(a) we show normalized \si{\micro}PL spectra measured at different temperatures. With increasing temperature the charged exciton peak broadens. The neutral exciton emission blue shifts (measured $\diff E_{\text{g}}/\diff T\sim\SI{3.1e-4}{\eV/\K}$ between \SI{40}{\K} and \SI{160}{\K}), which is consistent with earlier reports of the temperature dependence of the band gap of bulk black phosphorus ($\diff E_{\text{g}}/\diff T=\SI{2.8e-4}{\eV/\K}$ \cite{warschauer1963electrical} or \SI{2.33e-4}{\eV/\K} \cite{baba1991photoconduction}). The behavior of the emission energy of the excitonic peak is shown more in detail in Fig.~\ref{fig:Tdependence}(b). In bulk semiconductors, the variation of the band gap as a function of the temperature is direct consequence of the renormalization of the band gap via the electron-phonon interaction and of the thermal expansion of the lattice.\cite{cardona2005isotope}
In the framework of the two-oscillator model\cite{vina1984temperature,cardona2014temperature,dey2013origin,lian2006effects}, the band gap
$E_{\text{g}}$ is approximated by
\begin{equation}
E_{\text{g}}(T)=E_0+E_1\left(\frac{2}{\neper^{\frac{\hbar\omega_1}{\boltzmannconstant T}}-1}+1\right) +
E_2\left(\frac{2}{\neper^{\frac{\hbar\omega_2}{\boltzmannconstant T}}-1}+1\right), \nonumber
\end{equation}
where $E_0$ is the bare band gap (i.e.\ the low temperature band gap exhibited in the absence of zero point motion),
$E_1+E_2$ is the renormalization energy and $\hbar\omega_{1}=\SI{17.23}{\milli\eV}$ and
$\hbar\omega_2=\SI{52.82}{\milli\eV}$ denote the two oscillator energies, as extracted from the computed phonon density
of states of monolayer black phosphorus.\cite{aierken2015thermal} The neutral exciton emission energy is then given by
$E_{\text{g}}(T) - E_{\text{X}}$, where $E_{\text{X}}$ is the exciton binding energy. The solid curve shown in Fig.~\ref{fig:Tdependence}(b) is
obtained by fitting the experimental data, yielding $(E_0 - E_\text{X})=2.22\pm\SI{0.02}{\eV}$ (larger than the observed low
$T$ transition energy, suggesting the occurrence of band gap renormalization due to electron-phonon interaction),
$E_1=72.6\pm\SI{0.3}{\milli\eV}$, and $E_2=-297\pm\SI{3}{\milli\eV}$.


The integrated intensity of the excitonic transition as a function of the inverse substrate temperature $T^{-1}$
is shown in Fig.~\ref{fig:Tdependence}(c). With increasing temperature, the emission intensity decreases, owing to the
thermal activation of non-radiative recombination centers. To quantify the activation energy $E_{\text{A}}$, the
experimental data is fitted using $I(T)/I(T=0)=1/(1+a\neper^{-E_{\text{A}}/\boltzmannconstant
T})$,\cite{leroux1999temperature} where $a$ is related to the ratio of the radiative and non-radiative
lifetimes,\cite{fang2015investigation} to give $E_{\text{A}}=10\pm\SI{1}{\milli\eV}$, and $a=6.8\pm2$. The
significantly lower value of $E_{\text{A}}$ as compared to the computed exciton binding energy \cite{Castellanos14}
confirms that the decrease of the PL intensity at high $T$ is brought about by the thermal occupation of non radiative
recombination centers rather than the dissociation of the excitons.

In summary, we have performed a detailed investigation of the optical properties of monolayer black phosphorus
mechanically exfoliated in an Ar atmosphere. No significant degradation of the PL emission was induced by the laser
illumination, suggesting that this preparation method preserves the optical properties of the black phosphorus flakes.
The shift of the Raman modes measured on the exfoliated flakes with respect to the bulk black phosphorus confirms the
single layer character of the exfoliated black phosphorus flakes. The measured \si{\micro}PL spectra exhibit strong
emission lines, attributed to the recombination of neutral and charged excitonic complexes. It is not clear if the
enhanced emission energy, compared to the previously reported values,\cite{Liu14,Yang15,wang2015highly} is linked with
a larger intrinsic band gap or a reduced exciton binding energy. The excitonic nature of the observed PL was confirmed
by the observed polarization dependence and by the nearly linear increase of the emission intensity with the excitation
power. The increase in the emission energy with temperature was modeled with a two-oscillator model to account for the
temperature dependence of semiconductor band gap.

\begin{acknowledgments}
The authors gratefully acknowledge Baptiste Vignolle for his assistance with the glove box and for his careful proof
reading and Geert Rikken for providing the bulk black phosphorus. AAM acknowledges financial support from the French
foreign ministry. This work was partially supported by ANR JCJC project milliPICS, the R\'egion Midi-Pyr\'en\'ees under
contract MESR 13053031 and STCU project 5809.
\end{acknowledgments}

\bibliography{BlackPbib}

\begin{thebibliography}{49}%
\makeatletter
\providecommand \@ifxundefined [1]{%
 \@ifx{#1\undefined}
}%
\providecommand \@ifnum [1]{%
 \ifnum #1\expandafter \@firstoftwo
 \else \expandafter \@secondoftwo
 \fi
}%
\providecommand \@ifx [1]{%
 \ifx #1\expandafter \@firstoftwo
 \else \expandafter \@secondoftwo
 \fi
}%
\providecommand \natexlab [1]{#1}%
\providecommand \enquote  [1]{``#1''}%
\providecommand \bibnamefont  [1]{#1}%
\providecommand \bibfnamefont [1]{#1}%
\providecommand \citenamefont [1]{#1}%
\providecommand \href@noop [0]{\@secondoftwo}%
\providecommand \href [0]{\begingroup \@sanitize@url \@href}%
\providecommand \@href[1]{\@@startlink{#1}\@@href}%
\providecommand \@@href[1]{\endgroup#1\@@endlink}%
\providecommand \@sanitize@url [0]{\catcode `\\12\catcode `\$12\catcode
  `\&12\catcode `\#12\catcode `\^12\catcode `\_12\catcode `\%12\relax}%
\providecommand \@@startlink[1]{}%
\providecommand \@@endlink[0]{}%
\providecommand \url  [0]{\begingroup\@sanitize@url \@url }%
\providecommand \@url [1]{\endgroup\@href {#1}{\urlprefix }}%
\providecommand \urlprefix  [0]{URL }%
\providecommand \Eprint [0]{\href }%
\providecommand \doibase [0]{http://dx.doi.org/}%
\providecommand \selectlanguage [0]{\@gobble}%
\providecommand \bibinfo  [0]{\@secondoftwo}%
\providecommand \bibfield  [0]{\@secondoftwo}%
\providecommand \translation [1]{[#1]}%
\providecommand \BibitemOpen [0]{}%
\providecommand \bibitemStop [0]{}%
\providecommand \bibitemNoStop [0]{.\EOS\space}%
\providecommand \EOS [0]{\spacefactor3000\relax}%
\providecommand \BibitemShut  [1]{\csname bibitem#1\endcsname}%
\let\auto@bib@innerbib\@empty
\bibitem [{\citenamefont {Keyes}(1953)}]{Keyes53}%
  \BibitemOpen
  \bibfield  {author} {\bibinfo {author} {\bibfnamefont {R.~W.}\ \bibnamefont
  {Keyes}},\ }\href@noop {} {\bibfield  {journal} {\bibinfo  {journal}
  {Physical Review}\ }\textbf {\bibinfo {volume} {92}},\ \bibinfo {pages} {580}
  (\bibinfo {year} {1953})}\BibitemShut {NoStop}%
\bibitem [{\citenamefont {Warschauer}(1963)}]{warschauer1963electrical}%
  \BibitemOpen
  \bibfield  {author} {\bibinfo {author} {\bibfnamefont {D.}~\bibnamefont
  {Warschauer}},\ }\href@noop {} {\bibfield  {journal} {\bibinfo  {journal}
  {Journal of Applied Physics}\ }\textbf {\bibinfo {volume} {34}},\ \bibinfo
  {pages} {1853} (\bibinfo {year} {1963})}\BibitemShut {NoStop}%
\bibitem [{\citenamefont {Maruyama}\ \emph {et~al.}(1981)\citenamefont
  {Maruyama}, \citenamefont {Suzuki}, \citenamefont {Kobayashi},\ and\
  \citenamefont {Tanuma}}]{Maruyama81}%
  \BibitemOpen
  \bibfield  {author} {\bibinfo {author} {\bibfnamefont {Y.}~\bibnamefont
  {Maruyama}}, \bibinfo {author} {\bibfnamefont {S.}~\bibnamefont {Suzuki}},
  \bibinfo {author} {\bibfnamefont {K.}~\bibnamefont {Kobayashi}}, \ and\
  \bibinfo {author} {\bibfnamefont {S.}~\bibnamefont {Tanuma}},\ }\href@noop {}
  {\bibfield  {journal} {\bibinfo  {journal} {Physica B+C}\ }\textbf {\bibinfo
  {volume} {105}},\ \bibinfo {pages} {99 } (\bibinfo {year}
  {1981})}\BibitemShut {NoStop}%
\bibitem [{\citenamefont {Asahina}\ \emph {et~al.}(1983)\citenamefont
  {Asahina}, \citenamefont {Maruyama},\ and\ \citenamefont
  {Morita}}]{Asahina83}%
  \BibitemOpen
  \bibfield  {author} {\bibinfo {author} {\bibfnamefont {H.}~\bibnamefont
  {Asahina}}, \bibinfo {author} {\bibfnamefont {Y.}~\bibnamefont {Maruyama}}, \
  and\ \bibinfo {author} {\bibfnamefont {A.}~\bibnamefont {Morita}},\
  }\href@noop {} {\bibfield  {journal} {\bibinfo  {journal} {Physica B+C}\
  }\textbf {\bibinfo {volume} {117–118, Part 1}},\ \bibinfo {pages} {419 }
  (\bibinfo {year} {1983})}\BibitemShut {NoStop}%
\bibitem [{\citenamefont {Jamieson}(1963)}]{Jamieson63}%
  \BibitemOpen
  \bibfield  {author} {\bibinfo {author} {\bibfnamefont {J.~C.}\ \bibnamefont
  {Jamieson}},\ }\href@noop {} {\bibfield  {journal} {\bibinfo  {journal}
  {Science}\ }\textbf {\bibinfo {volume} {139}},\ \bibinfo {pages} {1291}
  (\bibinfo {year} {1963})}\BibitemShut {NoStop}%
\bibitem [{\citenamefont {Wittig}\ and\ \citenamefont
  {Matthias}(1968)}]{Wittig68}%
  \BibitemOpen
  \bibfield  {author} {\bibinfo {author} {\bibfnamefont {J.}~\bibnamefont
  {Wittig}}\ and\ \bibinfo {author} {\bibfnamefont {B.~T.}\ \bibnamefont
  {Matthias}},\ }\href@noop {} {\bibfield  {journal} {\bibinfo  {journal}
  {Science}\ }\textbf {\bibinfo {volume} {160}},\ \bibinfo {pages} {994}
  (\bibinfo {year} {1968})}\BibitemShut {NoStop}%
\bibitem [{\citenamefont {Sugai}\ and\ \citenamefont
  {Shirotani}(1985)}]{Sugai85}%
  \BibitemOpen
  \bibfield  {author} {\bibinfo {author} {\bibfnamefont {S.}~\bibnamefont
  {Sugai}}\ and\ \bibinfo {author} {\bibfnamefont {I.}~\bibnamefont
  {Shirotani}},\ }\href@noop {} {\bibfield  {journal} {\bibinfo  {journal}
  {Solid State Communications}\ }\textbf {\bibinfo {volume} {53}},\ \bibinfo
  {pages} {753 } (\bibinfo {year} {1985})}\BibitemShut {NoStop}%
\bibitem [{\citenamefont {Narita}\ \emph {et~al.}(1983)\citenamefont {Narita},
  \citenamefont {Terada}, \citenamefont {Mori}, \citenamefont {Muro},
  \citenamefont {Akahama},\ and\ \citenamefont {Endo}}]{Narita83}%
  \BibitemOpen
  \bibfield  {author} {\bibinfo {author} {\bibfnamefont {S.}~\bibnamefont
  {Narita}}, \bibinfo {author} {\bibfnamefont {S.}~\bibnamefont {Terada}},
  \bibinfo {author} {\bibfnamefont {S.}~\bibnamefont {Mori}}, \bibinfo {author}
  {\bibfnamefont {K.}~\bibnamefont {Muro}}, \bibinfo {author} {\bibfnamefont
  {Y.}~\bibnamefont {Akahama}}, \ and\ \bibinfo {author} {\bibfnamefont
  {S.}~\bibnamefont {Endo}},\ }\href@noop {} {\bibfield  {journal} {\bibinfo
  {journal} {Journal of the Physical Society of Japan}\ }\textbf {\bibinfo
  {volume} {52}},\ \bibinfo {pages} {3544} (\bibinfo {year}
  {1983})}\BibitemShut {NoStop}%
\bibitem [{\citenamefont {Takao}\ \emph {et~al.}(1981)\citenamefont {Takao},
  \citenamefont {Asahina},\ and\ \citenamefont {Morita}}]{Takao81}%
  \BibitemOpen
  \bibfield  {author} {\bibinfo {author} {\bibfnamefont {Y.}~\bibnamefont
  {Takao}}, \bibinfo {author} {\bibfnamefont {H.}~\bibnamefont {Asahina}}, \
  and\ \bibinfo {author} {\bibfnamefont {A.}~\bibnamefont {Morita}},\
  }\href@noop {} {\bibfield  {journal} {\bibinfo  {journal} {Journal of the
  Physical Society of Japan}\ }\textbf {\bibinfo {volume} {50}},\ \bibinfo
  {pages} {3362} (\bibinfo {year} {1981})}\BibitemShut {NoStop}%
\bibitem [{\citenamefont {Castellanos-Gomez}\ \emph {et~al.}(2014)\citenamefont
  {Castellanos-Gomez}, \citenamefont {Vicarelli}, \citenamefont {Prada},
  \citenamefont {Island}, \citenamefont {Narasimha-Acharya}, \citenamefont
  {Blanter}, \citenamefont {Groenendijk}, \citenamefont {Buscema},
  \citenamefont {Steele}, \citenamefont {Alvarez}, \citenamefont {Zandbergen},
  \citenamefont {Palacios},\ and\ \citenamefont {van~der
  Zant}}]{Castellanos14}%
  \BibitemOpen
  \bibfield  {author} {\bibinfo {author} {\bibfnamefont {A.}~\bibnamefont
  {Castellanos-Gomez}}, \bibinfo {author} {\bibfnamefont {L.}~\bibnamefont
  {Vicarelli}}, \bibinfo {author} {\bibfnamefont {E.}~\bibnamefont {Prada}},
  \bibinfo {author} {\bibfnamefont {J.~O.}\ \bibnamefont {Island}}, \bibinfo
  {author} {\bibfnamefont {K.~L.}\ \bibnamefont {Narasimha-Acharya}}, \bibinfo
  {author} {\bibfnamefont {S.~I.}\ \bibnamefont {Blanter}}, \bibinfo {author}
  {\bibfnamefont {D.~J.}\ \bibnamefont {Groenendijk}}, \bibinfo {author}
  {\bibfnamefont {M.}~\bibnamefont {Buscema}}, \bibinfo {author} {\bibfnamefont
  {G.~A.}\ \bibnamefont {Steele}}, \bibinfo {author} {\bibfnamefont {J.~V.}\
  \bibnamefont {Alvarez}}, \bibinfo {author} {\bibfnamefont {H.~W.}\
  \bibnamefont {Zandbergen}}, \bibinfo {author} {\bibfnamefont {J.~J.}\
  \bibnamefont {Palacios}}, \ and\ \bibinfo {author} {\bibfnamefont {H.~S.~J.}\
  \bibnamefont {van~der Zant}},\ }\href@noop {} {\bibfield  {journal} {\bibinfo
   {journal} {2D Materials}\ }\textbf {\bibinfo {volume} {1}},\ \bibinfo
  {pages} {025001} (\bibinfo {year} {2014})}\BibitemShut {NoStop}%
\bibitem [{\citenamefont {Tran}\ \emph {et~al.}(2014)\citenamefont {Tran},
  \citenamefont {Soklaski}, \citenamefont {Liang},\ and\ \citenamefont
  {Yang}}]{Tran14}%
  \BibitemOpen
  \bibfield  {author} {\bibinfo {author} {\bibfnamefont {V.}~\bibnamefont
  {Tran}}, \bibinfo {author} {\bibfnamefont {R.}~\bibnamefont {Soklaski}},
  \bibinfo {author} {\bibfnamefont {Y.}~\bibnamefont {Liang}}, \ and\ \bibinfo
  {author} {\bibfnamefont {L.}~\bibnamefont {Yang}},\ }\href@noop {} {\bibfield
   {journal} {\bibinfo  {journal} {Physical Review B}\ }\textbf {\bibinfo
  {volume} {89}},\ \bibinfo {pages} {235319} (\bibinfo {year}
  {2014})}\BibitemShut {NoStop}%
\bibitem [{\citenamefont {Ling}\ \emph {et~al.}(2015)\citenamefont {Ling},
  \citenamefont {Wang}, \citenamefont {Huang}, \citenamefont {Xia},\ and\
  \citenamefont {Dresselhaus}}]{ling2015}%
  \BibitemOpen
  \bibfield  {author} {\bibinfo {author} {\bibfnamefont {X.}~\bibnamefont
  {Ling}}, \bibinfo {author} {\bibfnamefont {H.}~\bibnamefont {Wang}}, \bibinfo
  {author} {\bibfnamefont {S.}~\bibnamefont {Huang}}, \bibinfo {author}
  {\bibfnamefont {F.}~\bibnamefont {Xia}}, \ and\ \bibinfo {author}
  {\bibfnamefont {M.~S.}\ \bibnamefont {Dresselhaus}},\ }\href@noop {}
  {\bibfield  {journal} {\bibinfo  {journal} {Proceedings of the National
  Academy of Sciences}\ }\textbf {\bibinfo {volume} {112}},\ \bibinfo {pages}
  {4523} (\bibinfo {year} {2015})}\BibitemShut {NoStop}%
\bibitem [{\citenamefont {Novoselov}\ \emph {et~al.}(2005)\citenamefont
  {Novoselov}, \citenamefont {Geim}, \citenamefont {Morozow}, \citenamefont
  {Jiang}, \citenamefont {Katsnelson}, \citenamefont {Grigorieva},
  \citenamefont {Dubonos},\ and\ \citenamefont {Firsov}}]{Novoselov05}%
  \BibitemOpen
  \bibfield  {author} {\bibinfo {author} {\bibfnamefont {K.~S.}\ \bibnamefont
  {Novoselov}}, \bibinfo {author} {\bibfnamefont {A.~K.}\ \bibnamefont {Geim}},
  \bibinfo {author} {\bibfnamefont {S.~V.}\ \bibnamefont {Morozow}}, \bibinfo
  {author} {\bibfnamefont {D.}~\bibnamefont {Jiang}}, \bibinfo {author}
  {\bibfnamefont {M.~I.}\ \bibnamefont {Katsnelson}}, \bibinfo {author}
  {\bibfnamefont {I.~V.}\ \bibnamefont {Grigorieva}}, \bibinfo {author}
  {\bibfnamefont {S.~V.}\ \bibnamefont {Dubonos}}, \ and\ \bibinfo {author}
  {\bibfnamefont {A.~A.}\ \bibnamefont {Firsov}},\ }\href@noop {} {\bibfield
  {journal} {\bibinfo  {journal} {Nature}\ }\textbf {\bibinfo {volume} {438}},\
  \bibinfo {pages} {197} (\bibinfo {year} {2005})}\BibitemShut {NoStop}%
\bibitem [{\citenamefont {Wang}\ \emph {et~al.}(2012)\citenamefont {Wang},
  \citenamefont {Kalantar-Zadeh}, \citenamefont {Kis}, \citenamefont
  {Coleman},\ and\ \citenamefont {Strano}}]{Wang12}%
  \BibitemOpen
  \bibfield  {author} {\bibinfo {author} {\bibfnamefont {Q.~H.}\ \bibnamefont
  {Wang}}, \bibinfo {author} {\bibfnamefont {K.}~\bibnamefont
  {Kalantar-Zadeh}}, \bibinfo {author} {\bibfnamefont {A.}~\bibnamefont {Kis}},
  \bibinfo {author} {\bibfnamefont {J.~N.}\ \bibnamefont {Coleman}}, \ and\
  \bibinfo {author} {\bibfnamefont {M.~S.}\ \bibnamefont {Strano}},\
  }\href@noop {} {\bibfield  {journal} {\bibinfo  {journal} {Nature
  Nanotechnology}\ }\textbf {\bibinfo {volume} {7}},\ \bibinfo {pages} {699}
  (\bibinfo {year} {2012})}\BibitemShut {NoStop}%
\bibitem [{\citenamefont {Low}\ \emph {et~al.}(2014)\citenamefont {Low},
  \citenamefont {Rodin}, \citenamefont {Carvalho}, \citenamefont {Jiang},
  \citenamefont {Wang}, \citenamefont {Xia},\ and\ \citenamefont
  {Castro~Neto}}]{Low14}%
  \BibitemOpen
  \bibfield  {author} {\bibinfo {author} {\bibfnamefont {T.}~\bibnamefont
  {Low}}, \bibinfo {author} {\bibfnamefont {A.~S.}\ \bibnamefont {Rodin}},
  \bibinfo {author} {\bibfnamefont {A.}~\bibnamefont {Carvalho}}, \bibinfo
  {author} {\bibfnamefont {Y.}~\bibnamefont {Jiang}}, \bibinfo {author}
  {\bibfnamefont {H.}~\bibnamefont {Wang}}, \bibinfo {author} {\bibfnamefont
  {F.}~\bibnamefont {Xia}}, \ and\ \bibinfo {author} {\bibfnamefont {A.~H.}\
  \bibnamefont {Castro~Neto}},\ }\href@noop {} {\bibfield  {journal} {\bibinfo
  {journal} {Physical Review B}\ }\textbf {\bibinfo {volume} {90}},\ \bibinfo
  {pages} {075434} (\bibinfo {year} {2014})}\BibitemShut {NoStop}%
\bibitem [{\citenamefont {Xia}\ \emph {et~al.}(2014)\citenamefont {Xia},
  \citenamefont {Wang},\ and\ \citenamefont {Jia}}]{xia14}%
  \BibitemOpen
  \bibfield  {author} {\bibinfo {author} {\bibfnamefont {F.}~\bibnamefont
  {Xia}}, \bibinfo {author} {\bibfnamefont {H.}~\bibnamefont {Wang}}, \ and\
  \bibinfo {author} {\bibfnamefont {Y.}~\bibnamefont {Jia}},\ }\href@noop {}
  {\bibfield  {journal} {\bibinfo  {journal} {Nature Communications}\ }\textbf
  {\bibinfo {volume} {5}},\ \bibinfo {pages} {4458} (\bibinfo {year}
  {2014})}\BibitemShut {NoStop}%
\bibitem [{\citenamefont {Qiao}\ \emph {et~al.}(2014)\citenamefont {Qiao},
  \citenamefont {Kong}, \citenamefont {Hu}, \citenamefont {Yang},\ and\
  \citenamefont {Ji}}]{Qiao14}%
  \BibitemOpen
  \bibfield  {author} {\bibinfo {author} {\bibfnamefont {J.}~\bibnamefont
  {Qiao}}, \bibinfo {author} {\bibfnamefont {X.}~\bibnamefont {Kong}}, \bibinfo
  {author} {\bibfnamefont {Z.-X.}\ \bibnamefont {Hu}}, \bibinfo {author}
  {\bibfnamefont {F.}~\bibnamefont {Yang}}, \ and\ \bibinfo {author}
  {\bibfnamefont {W.}~\bibnamefont {Ji}},\ }\href@noop {} {\bibfield  {journal}
  {\bibinfo  {journal} {Nature Communications}\ }\textbf {\bibinfo {volume}
  {5}},\ \bibinfo {pages} {4475} (\bibinfo {year} {2014})}\BibitemShut
  {NoStop}%
\bibitem [{\citenamefont {Li}\ \emph {et~al.}(2014)\citenamefont {Li},
  \citenamefont {Yu}, \citenamefont {Ye}, \citenamefont {Ge}, \citenamefont
  {Ou}, \citenamefont {Wu}, \citenamefont {Feng}, \citenamefont {Chen},\ and\
  \citenamefont {Zhang}}]{li14}%
  \BibitemOpen
  \bibfield  {author} {\bibinfo {author} {\bibfnamefont {L.}~\bibnamefont
  {Li}}, \bibinfo {author} {\bibfnamefont {Y.}~\bibnamefont {Yu}}, \bibinfo
  {author} {\bibfnamefont {G.~J.}\ \bibnamefont {Ye}}, \bibinfo {author}
  {\bibfnamefont {Q.}~\bibnamefont {Ge}}, \bibinfo {author} {\bibfnamefont
  {X.}~\bibnamefont {Ou}}, \bibinfo {author} {\bibfnamefont {H.}~\bibnamefont
  {Wu}}, \bibinfo {author} {\bibfnamefont {D.}~\bibnamefont {Feng}}, \bibinfo
  {author} {\bibfnamefont {X.~H.}\ \bibnamefont {Chen}}, \ and\ \bibinfo
  {author} {\bibfnamefont {Y.}~\bibnamefont {Zhang}},\ }\href@noop {}
  {\bibfield  {journal} {\bibinfo  {journal} {Nature Nanotechnology}\ }\textbf
  {\bibinfo {volume} {9}},\ \bibinfo {pages} {372} (\bibinfo {year}
  {2014})}\BibitemShut {NoStop}%
\bibitem [{\citenamefont {Liu}\ \emph {et~al.}(2014)\citenamefont {Liu},
  \citenamefont {Neal}, \citenamefont {Zhu}, \citenamefont {Luo}, \citenamefont
  {Xu}, \citenamefont {Tom\'{a}nek},\ and\ \citenamefont {Ye}}]{Liu14}%
  \BibitemOpen
  \bibfield  {author} {\bibinfo {author} {\bibfnamefont {H.}~\bibnamefont
  {Liu}}, \bibinfo {author} {\bibfnamefont {A.~T.}\ \bibnamefont {Neal}},
  \bibinfo {author} {\bibfnamefont {Z.}~\bibnamefont {Zhu}}, \bibinfo {author}
  {\bibfnamefont {Z.}~\bibnamefont {Luo}}, \bibinfo {author} {\bibfnamefont
  {X.}~\bibnamefont {Xu}}, \bibinfo {author} {\bibfnamefont {D.}~\bibnamefont
  {Tom\'{a}nek}}, \ and\ \bibinfo {author} {\bibfnamefont {P.~D.}\ \bibnamefont
  {Ye}},\ }\href@noop {} {\bibfield  {journal} {\bibinfo  {journal} {ACS Nano}\
  }\textbf {\bibinfo {volume} {8}},\ \bibinfo {pages} {4033} (\bibinfo {year}
  {2014})}\BibitemShut {NoStop}%
\bibitem [{\citenamefont {Buscema}\ \emph {et~al.}(2014)\citenamefont
  {Buscema}, \citenamefont {Groenendijk}, \citenamefont {Blanter},
  \citenamefont {Steele}, \citenamefont {van~der Zant},\ and\ \citenamefont
  {Castellanos-Gomez}}]{Buscema15}%
  \BibitemOpen
  \bibfield  {author} {\bibinfo {author} {\bibfnamefont {M.}~\bibnamefont
  {Buscema}}, \bibinfo {author} {\bibfnamefont {D.~J.}\ \bibnamefont
  {Groenendijk}}, \bibinfo {author} {\bibfnamefont {S.~I.}\ \bibnamefont
  {Blanter}}, \bibinfo {author} {\bibfnamefont {G.~A.}\ \bibnamefont {Steele}},
  \bibinfo {author} {\bibfnamefont {H.~S.~J.}\ \bibnamefont {van~der Zant}}, \
  and\ \bibinfo {author} {\bibfnamefont {A.}~\bibnamefont
  {Castellanos-Gomez}},\ }\href@noop {} {\bibfield  {journal} {\bibinfo
  {journal} {Nano Letters}\ }\textbf {\bibinfo {volume} {14}},\ \bibinfo
  {pages} {3347} (\bibinfo {year} {2014})}\BibitemShut {NoStop}%
\bibitem [{\citenamefont {Koenig}\ \emph {et~al.}(2014)\citenamefont {Koenig},
  \citenamefont {Doganov}, \citenamefont {Schmidt}, \citenamefont
  {Castro~Neto},\ and\ \citenamefont {\"{O}zyilmaz}}]{Koenig14}%
  \BibitemOpen
  \bibfield  {author} {\bibinfo {author} {\bibfnamefont {S.~P.}\ \bibnamefont
  {Koenig}}, \bibinfo {author} {\bibfnamefont {R.~A.}\ \bibnamefont {Doganov}},
  \bibinfo {author} {\bibfnamefont {H.}~\bibnamefont {Schmidt}}, \bibinfo
  {author} {\bibfnamefont {A.~H.}\ \bibnamefont {Castro~Neto}}, \ and\ \bibinfo
  {author} {\bibfnamefont {B.}~\bibnamefont {\"{O}zyilmaz}},\ }\href@noop {}
  {\bibfield  {journal} {\bibinfo  {journal} {Applied Physics Letters}\
  }\textbf {\bibinfo {volume} {104}},\ \bibinfo {pages} {103106} (\bibinfo
  {year} {2014})}\BibitemShut {NoStop}%
\bibitem [{\citenamefont {Gan}\ \emph {et~al.}(2015)\citenamefont {Gan},
  \citenamefont {Sun}, \citenamefont {Wu}, \citenamefont {Meng}, \citenamefont
  {Shen},\ and\ \citenamefont {Chu}}]{Gan15}%
  \BibitemOpen
  \bibfield  {author} {\bibinfo {author} {\bibfnamefont {Z.~X.}\ \bibnamefont
  {Gan}}, \bibinfo {author} {\bibfnamefont {L.~L.}\ \bibnamefont {Sun}},
  \bibinfo {author} {\bibfnamefont {X.~L.}\ \bibnamefont {Wu}}, \bibinfo
  {author} {\bibfnamefont {M.}~\bibnamefont {Meng}}, \bibinfo {author}
  {\bibfnamefont {J.~C.}\ \bibnamefont {Shen}}, \ and\ \bibinfo {author}
  {\bibfnamefont {P.~K.}\ \bibnamefont {Chu}},\ }\href@noop {} {\bibfield
  {journal} {\bibinfo  {journal} {Applied Physics Letters}\ }\textbf {\bibinfo
  {volume} {107}},\ \bibinfo {pages} {021901} (\bibinfo {year}
  {2015})}\BibitemShut {NoStop}%
\bibitem [{\citenamefont {Zhang}\ \emph {et~al.}(2014)\citenamefont {Zhang},
  \citenamefont {Yang}, \citenamefont {Xu}, \citenamefont {Wang}, \citenamefont
  {Li}, \citenamefont {Ghufran}, \citenamefont {Zhang}, \citenamefont {Yu},
  \citenamefont {Zhang}, \citenamefont {Qin},\ and\ \citenamefont
  {Lu}}]{Zhang14}%
  \BibitemOpen
  \bibfield  {author} {\bibinfo {author} {\bibfnamefont {S.}~\bibnamefont
  {Zhang}}, \bibinfo {author} {\bibfnamefont {J.}~\bibnamefont {Yang}},
  \bibinfo {author} {\bibfnamefont {R.}~\bibnamefont {Xu}}, \bibinfo {author}
  {\bibfnamefont {F.}~\bibnamefont {Wang}}, \bibinfo {author} {\bibfnamefont
  {W.}~\bibnamefont {Li}}, \bibinfo {author} {\bibfnamefont {M.}~\bibnamefont
  {Ghufran}}, \bibinfo {author} {\bibfnamefont {Y.-W.}\ \bibnamefont {Zhang}},
  \bibinfo {author} {\bibfnamefont {Z.}~\bibnamefont {Yu}}, \bibinfo {author}
  {\bibfnamefont {G.}~\bibnamefont {Zhang}}, \bibinfo {author} {\bibfnamefont
  {Q.}~\bibnamefont {Qin}}, \ and\ \bibinfo {author} {\bibfnamefont
  {Y.}~\bibnamefont {Lu}},\ }\href@noop {} {\bibfield  {journal} {\bibinfo
  {journal} {ACS Nano}\ }\textbf {\bibinfo {volume} {8}},\ \bibinfo {pages}
  {9590} (\bibinfo {year} {2014})}\BibitemShut {NoStop}%
\bibitem [{\citenamefont {Wang}\ \emph {et~al.}(2015)\citenamefont {Wang},
  \citenamefont {Jones}, \citenamefont {Seyler}, \citenamefont {Tran},
  \citenamefont {Jia}, \citenamefont {Zhao}, \citenamefont {Wang},
  \citenamefont {Yang}, \citenamefont {Xu},\ and\ \citenamefont
  {Xia}}]{wang2015highly}%
  \BibitemOpen
  \bibfield  {author} {\bibinfo {author} {\bibfnamefont {X.}~\bibnamefont
  {Wang}}, \bibinfo {author} {\bibfnamefont {A.~M.}\ \bibnamefont {Jones}},
  \bibinfo {author} {\bibfnamefont {K.~L.}\ \bibnamefont {Seyler}}, \bibinfo
  {author} {\bibfnamefont {V.}~\bibnamefont {Tran}}, \bibinfo {author}
  {\bibfnamefont {Y.}~\bibnamefont {Jia}}, \bibinfo {author} {\bibfnamefont
  {H.}~\bibnamefont {Zhao}}, \bibinfo {author} {\bibfnamefont {H.}~\bibnamefont
  {Wang}}, \bibinfo {author} {\bibfnamefont {L.}~\bibnamefont {Yang}}, \bibinfo
  {author} {\bibfnamefont {X.}~\bibnamefont {Xu}}, \ and\ \bibinfo {author}
  {\bibfnamefont {F.}~\bibnamefont {Xia}},\ }\href@noop {} {\bibfield
  {journal} {\bibinfo  {journal} {Nature Nanotechnology}\ }\textbf {\bibinfo
  {volume} {10}},\ \bibinfo {pages} {517} (\bibinfo {year} {2015})}\BibitemShut
  {NoStop}%
\bibitem [{\citenamefont {Yang}\ \emph {et~al.}(2015)\citenamefont {Yang},
  \citenamefont {Xu}, \citenamefont {Pei}, \citenamefont {Myint}, \citenamefont
  {Wang}, \citenamefont {Wang}, \citenamefont {Zhang}, \citenamefont {Yu},\
  and\ \citenamefont {Lu}}]{Yang15}%
  \BibitemOpen
  \bibfield  {author} {\bibinfo {author} {\bibfnamefont {J.}~\bibnamefont
  {Yang}}, \bibinfo {author} {\bibfnamefont {R.}~\bibnamefont {Xu}}, \bibinfo
  {author} {\bibfnamefont {J.}~\bibnamefont {Pei}}, \bibinfo {author}
  {\bibfnamefont {Y.~W.}\ \bibnamefont {Myint}}, \bibinfo {author}
  {\bibfnamefont {F.}~\bibnamefont {Wang}}, \bibinfo {author} {\bibfnamefont
  {Z.}~\bibnamefont {Wang}}, \bibinfo {author} {\bibfnamefont {S.}~\bibnamefont
  {Zhang}}, \bibinfo {author} {\bibfnamefont {Z.}~\bibnamefont {Yu}}, \ and\
  \bibinfo {author} {\bibfnamefont {Y.}~\bibnamefont {Lu}},\ }\href@noop {}
  {\bibfield  {journal} {\bibinfo  {journal} {{Light-Science \& Applications}}\
  }\textbf {\bibinfo {volume} {{4}}},\ \bibinfo {pages} {e312} (\bibinfo {year}
  {{2015}})}\BibitemShut {NoStop}%
\bibitem [{\citenamefont {Liang}\ \emph {et~al.}(2014)\citenamefont {Liang},
  \citenamefont {Wang}, \citenamefont {Lin}, \citenamefont {Sumpter},
  \citenamefont {Meunier},\ and\ \citenamefont {Pan}}]{liang2014electronic}%
  \BibitemOpen
  \bibfield  {author} {\bibinfo {author} {\bibfnamefont {L.}~\bibnamefont
  {Liang}}, \bibinfo {author} {\bibfnamefont {J.}~\bibnamefont {Wang}},
  \bibinfo {author} {\bibfnamefont {W.}~\bibnamefont {Lin}}, \bibinfo {author}
  {\bibfnamefont {B.~G.}\ \bibnamefont {Sumpter}}, \bibinfo {author}
  {\bibfnamefont {V.}~\bibnamefont {Meunier}}, \ and\ \bibinfo {author}
  {\bibfnamefont {M.}~\bibnamefont {Pan}},\ }\href@noop {} {\bibfield
  {journal} {\bibinfo  {journal} {Nano Letters}\ }\textbf {\bibinfo {volume}
  {14}},\ \bibinfo {pages} {6400} (\bibinfo {year} {2014})}\BibitemShut
  {NoStop}%
\bibitem [{\citenamefont {Woomer}\ \emph {et~al.}(2015)\citenamefont {Woomer},
  \citenamefont {Farnsworth}, \citenamefont {Hu}, \citenamefont {Wells},
  \citenamefont {Donley},\ and\ \citenamefont {Warren}}]{Woomer15}%
  \BibitemOpen
  \bibfield  {author} {\bibinfo {author} {\bibfnamefont {A.~H.}\ \bibnamefont
  {Woomer}}, \bibinfo {author} {\bibfnamefont {T.~W.}\ \bibnamefont
  {Farnsworth}}, \bibinfo {author} {\bibfnamefont {J.}~\bibnamefont {Hu}},
  \bibinfo {author} {\bibfnamefont {R.~A.}\ \bibnamefont {Wells}}, \bibinfo
  {author} {\bibfnamefont {C.~L.}\ \bibnamefont {Donley}}, \ and\ \bibinfo
  {author} {\bibfnamefont {S.~C.}\ \bibnamefont {Warren}},\ }\href@noop {}
  {\bibfield  {journal} {\bibinfo  {journal} {ACS Nano}\ }\textbf {\bibinfo
  {volume} {9}},\ \bibinfo {pages} {8869} (\bibinfo {year} {2015})}\BibitemShut
  {NoStop}%
\bibitem [{\citenamefont {Lee}\ \emph {et~al.}(2010)\citenamefont {Lee},
  \citenamefont {Yan}, \citenamefont {Brus}, \citenamefont {Heinz},
  \citenamefont {Hone},\ and\ \citenamefont {Ryu}}]{Lee10}%
  \BibitemOpen
  \bibfield  {author} {\bibinfo {author} {\bibfnamefont {C.}~\bibnamefont
  {Lee}}, \bibinfo {author} {\bibfnamefont {H.}~\bibnamefont {Yan}}, \bibinfo
  {author} {\bibfnamefont {L.~E.}\ \bibnamefont {Brus}}, \bibinfo {author}
  {\bibfnamefont {T.~F.}\ \bibnamefont {Heinz}}, \bibinfo {author}
  {\bibfnamefont {J.}~\bibnamefont {Hone}}, \ and\ \bibinfo {author}
  {\bibfnamefont {S.}~\bibnamefont {Ryu}},\ }\href@noop {} {\bibfield
  {journal} {\bibinfo  {journal} {ACS Nano}\ }\textbf {\bibinfo {volume} {4}},\
  \bibinfo {pages} {2695} (\bibinfo {year} {2010})}\BibitemShut {NoStop}%
\bibitem [{\citenamefont {Gutiérrez}\ \emph {et~al.}(2013)\citenamefont
  {Gutiérrez}, \citenamefont {Perea-López}, \citenamefont {Elías},
  \citenamefont {Berkdemir}, \citenamefont {Wang}, \citenamefont {Lv},
  \citenamefont {López-Urías}, \citenamefont {Crespi}, \citenamefont
  {Terrones},\ and\ \citenamefont {Terrones}}]{Gutierrez13}%
  \BibitemOpen
  \bibfield  {author} {\bibinfo {author} {\bibfnamefont {H.~R.}\ \bibnamefont
  {Gutiérrez}}, \bibinfo {author} {\bibfnamefont {N.}~\bibnamefont
  {Perea-López}}, \bibinfo {author} {\bibfnamefont {A.~L.}\ \bibnamefont
  {Elías}}, \bibinfo {author} {\bibfnamefont {A.}~\bibnamefont {Berkdemir}},
  \bibinfo {author} {\bibfnamefont {B.}~\bibnamefont {Wang}}, \bibinfo {author}
  {\bibfnamefont {R.}~\bibnamefont {Lv}}, \bibinfo {author} {\bibfnamefont
  {F.}~\bibnamefont {López-Urías}}, \bibinfo {author} {\bibfnamefont {V.~H.}\
  \bibnamefont {Crespi}}, \bibinfo {author} {\bibfnamefont {H.}~\bibnamefont
  {Terrones}}, \ and\ \bibinfo {author} {\bibfnamefont {M.}~\bibnamefont
  {Terrones}},\ }\href@noop {} {\bibfield  {journal} {\bibinfo  {journal} {Nano
  Letters}\ }\textbf {\bibinfo {volume} {13}},\ \bibinfo {pages} {3447}
  (\bibinfo {year} {2013})}\BibitemShut {NoStop}%
\bibitem [{\citenamefont {Li}\ \emph {et~al.}(2012{\natexlab{a}})\citenamefont
  {Li}, \citenamefont {Zhang}, \citenamefont {Yap}, \citenamefont {Tay},
  \citenamefont {Edwin}, \citenamefont {Olivier},\ and\ \citenamefont
  {Baillargeat}}]{Li12a}%
  \BibitemOpen
  \bibfield  {author} {\bibinfo {author} {\bibfnamefont {H.}~\bibnamefont
  {Li}}, \bibinfo {author} {\bibfnamefont {Q.}~\bibnamefont {Zhang}}, \bibinfo
  {author} {\bibfnamefont {C.~C.~R.}\ \bibnamefont {Yap}}, \bibinfo {author}
  {\bibfnamefont {B.~K.}\ \bibnamefont {Tay}}, \bibinfo {author} {\bibfnamefont
  {T.~H.~T.}\ \bibnamefont {Edwin}}, \bibinfo {author} {\bibfnamefont
  {A.}~\bibnamefont {Olivier}}, \ and\ \bibinfo {author} {\bibfnamefont
  {D.}~\bibnamefont {Baillargeat}},\ }\href@noop {} {\bibfield  {journal}
  {\bibinfo  {journal} {Advanced Functional Materials}\ }\textbf {\bibinfo
  {volume} {22}},\ \bibinfo {pages} {1385} (\bibinfo {year}
  {2012}{\natexlab{a}})}\BibitemShut {NoStop}%
\bibitem [{\citenamefont {Li}\ \emph {et~al.}(2012{\natexlab{b}})\citenamefont
  {Li}, \citenamefont {Miyazaki}, \citenamefont {Song}, \citenamefont
  {Kuramochi}, \citenamefont {Nakaharai},\ and\ \citenamefont
  {Tsukagoshi}}]{Li12}%
  \BibitemOpen
  \bibfield  {author} {\bibinfo {author} {\bibfnamefont {S.-L.}\ \bibnamefont
  {Li}}, \bibinfo {author} {\bibfnamefont {H.}~\bibnamefont {Miyazaki}},
  \bibinfo {author} {\bibfnamefont {H.}~\bibnamefont {Song}}, \bibinfo {author}
  {\bibfnamefont {H.}~\bibnamefont {Kuramochi}}, \bibinfo {author}
  {\bibfnamefont {S.}~\bibnamefont {Nakaharai}}, \ and\ \bibinfo {author}
  {\bibfnamefont {K.}~\bibnamefont {Tsukagoshi}},\ }\href@noop {} {\bibfield
  {journal} {\bibinfo  {journal} {ACS Nano}\ }\textbf {\bibinfo {volume} {6}},\
  \bibinfo {pages} {7381} (\bibinfo {year} {2012}{\natexlab{b}})}\BibitemShut
  {NoStop}%
\bibitem [{\citenamefont {Ferrari}\ \emph {et~al.}(2006)\citenamefont
  {Ferrari}, \citenamefont {Meyer}, \citenamefont {Scardaci}, \citenamefont
  {Casiraghi}, \citenamefont {Lazzeri}, \citenamefont {Mauri}, \citenamefont
  {Piscanec}, \citenamefont {Jiang}, \citenamefont {Novoselov}, \citenamefont
  {Roth},\ and\ \citenamefont {Geim}}]{Ferrari06}%
  \BibitemOpen
  \bibfield  {author} {\bibinfo {author} {\bibfnamefont {A.~C.}\ \bibnamefont
  {Ferrari}}, \bibinfo {author} {\bibfnamefont {J.~C.}\ \bibnamefont {Meyer}},
  \bibinfo {author} {\bibfnamefont {V.}~\bibnamefont {Scardaci}}, \bibinfo
  {author} {\bibfnamefont {C.}~\bibnamefont {Casiraghi}}, \bibinfo {author}
  {\bibfnamefont {M.}~\bibnamefont {Lazzeri}}, \bibinfo {author} {\bibfnamefont
  {F.}~\bibnamefont {Mauri}}, \bibinfo {author} {\bibfnamefont
  {S.}~\bibnamefont {Piscanec}}, \bibinfo {author} {\bibfnamefont
  {D.}~\bibnamefont {Jiang}}, \bibinfo {author} {\bibfnamefont {K.~S.}\
  \bibnamefont {Novoselov}}, \bibinfo {author} {\bibfnamefont {S.}~\bibnamefont
  {Roth}}, \ and\ \bibinfo {author} {\bibfnamefont {A.~K.}\ \bibnamefont
  {Geim}},\ }\href@noop {} {\bibfield  {journal} {\bibinfo  {journal} {Physical
  Review Letters}\ }\textbf {\bibinfo {volume} {97}},\ \bibinfo {pages}
  {187401} (\bibinfo {year} {2006})}\BibitemShut {NoStop}%
\bibitem [{\citenamefont {Gupta}\ \emph {et~al.}(2006)\citenamefont {Gupta},
  \citenamefont {Chen}, \citenamefont {Joshi}, \citenamefont {Tadigadapa},\
  and\ \citenamefont {Eklund}}]{Gupta06}%
  \BibitemOpen
  \bibfield  {author} {\bibinfo {author} {\bibfnamefont {A.}~\bibnamefont
  {Gupta}}, \bibinfo {author} {\bibfnamefont {G.}~\bibnamefont {Chen}},
  \bibinfo {author} {\bibfnamefont {P.}~\bibnamefont {Joshi}}, \bibinfo
  {author} {\bibfnamefont {S.}~\bibnamefont {Tadigadapa}}, \ and\ \bibinfo
  {author} {\bibnamefont {Eklund}},\ }\href@noop {} {\bibfield  {journal}
  {\bibinfo  {journal} {Nano Letters}\ }\textbf {\bibinfo {volume} {6}},\
  \bibinfo {pages} {2667} (\bibinfo {year} {2006})}\BibitemShut {NoStop}%
\bibitem [{\citenamefont {Sugai}\ \emph {et~al.}(1981)\citenamefont {Sugai},
  \citenamefont {Ueda},\ and\ \citenamefont {Murase}}]{Sugai81}%
  \BibitemOpen
  \bibfield  {author} {\bibinfo {author} {\bibfnamefont {S.}~\bibnamefont
  {Sugai}}, \bibinfo {author} {\bibfnamefont {T.}~\bibnamefont {Ueda}}, \ and\
  \bibinfo {author} {\bibfnamefont {K.}~\bibnamefont {Murase}},\ }\href@noop {}
  {\bibfield  {journal} {\bibinfo  {journal} {Journal of the Physical Society
  of Japan}\ }\textbf {\bibinfo {volume} {50}},\ \bibinfo {pages} {3356}
  (\bibinfo {year} {1981})}\BibitemShut {NoStop}%
\bibitem [{\citenamefont {Favron}\ \emph {et~al.}(2015)\citenamefont {Favron},
  \citenamefont {Gaufres}, \citenamefont {Fossard}, \citenamefont
  {Phaneuf-L'Heureux}, \citenamefont {Tang}, \citenamefont {Levesque},
  \citenamefont {Loiseau}, \citenamefont {Leonelli}, \citenamefont
  {Francoeur},\ and\ \citenamefont {Martel}}]{Favron15}%
  \BibitemOpen
  \bibfield  {author} {\bibinfo {author} {\bibfnamefont {A.}~\bibnamefont
  {Favron}}, \bibinfo {author} {\bibfnamefont {E.}~\bibnamefont {Gaufres}},
  \bibinfo {author} {\bibfnamefont {F.}~\bibnamefont {Fossard}}, \bibinfo
  {author} {\bibfnamefont {A.-L.}\ \bibnamefont {Phaneuf-L'Heureux}}, \bibinfo
  {author} {\bibfnamefont {N.~Y.-W.}\ \bibnamefont {Tang}}, \bibinfo {author}
  {\bibfnamefont {P.~L.}\ \bibnamefont {Levesque}}, \bibinfo {author}
  {\bibfnamefont {A.}~\bibnamefont {Loiseau}}, \bibinfo {author} {\bibfnamefont
  {R.}~\bibnamefont {Leonelli}}, \bibinfo {author} {\bibfnamefont
  {S.}~\bibnamefont {Francoeur}}, \ and\ \bibinfo {author} {\bibfnamefont
  {R.}~\bibnamefont {Martel}},\ }\href@noop {} {\bibfield  {journal} {\bibinfo
  {journal} {{Nature Materials}}\ }\textbf {\bibinfo {volume} {{14}}},\
  \bibinfo {pages} {{826}} (\bibinfo {year} {2015})}\BibitemShut {NoStop}%
\bibitem [{\citenamefont {Lu}\ \emph {et~al.}(2014)\citenamefont {Lu},
  \citenamefont {Nan}, \citenamefont {Hong}, \citenamefont {Chen},
  \citenamefont {Zhu}, \citenamefont {Liang}, \citenamefont {Ma}, \citenamefont
  {Ni}, \citenamefont {Jin},\ and\ \citenamefont {Zhang}}]{lu2014plasma}%
  \BibitemOpen
  \bibfield  {author} {\bibinfo {author} {\bibfnamefont {W.}~\bibnamefont
  {Lu}}, \bibinfo {author} {\bibfnamefont {H.}~\bibnamefont {Nan}}, \bibinfo
  {author} {\bibfnamefont {J.}~\bibnamefont {Hong}}, \bibinfo {author}
  {\bibfnamefont {Y.}~\bibnamefont {Chen}}, \bibinfo {author} {\bibfnamefont
  {C.}~\bibnamefont {Zhu}}, \bibinfo {author} {\bibfnamefont {Z.}~\bibnamefont
  {Liang}}, \bibinfo {author} {\bibfnamefont {X.}~\bibnamefont {Ma}}, \bibinfo
  {author} {\bibfnamefont {Z.}~\bibnamefont {Ni}}, \bibinfo {author}
  {\bibfnamefont {C.}~\bibnamefont {Jin}}, \ and\ \bibinfo {author}
  {\bibfnamefont {Z.}~\bibnamefont {Zhang}},\ }\href@noop {} {\bibfield
  {journal} {\bibinfo  {journal} {Nano Research}\ }\textbf {\bibinfo {volume}
  {7}},\ \bibinfo {pages} {853} (\bibinfo {year} {2014})}\BibitemShut {NoStop}%
\bibitem [{\citenamefont {Luo}\ \emph {et~al.}(2015)\citenamefont {Luo},
  \citenamefont {Lu}, \citenamefont {Koon}, \citenamefont {Neto}, \citenamefont
  {\"{O}zyilmaz}, \citenamefont {Xiong},\ and\ \citenamefont {Quek}}]{Luo15}%
  \BibitemOpen
  \bibfield  {author} {\bibinfo {author} {\bibfnamefont {X.}~\bibnamefont
  {Luo}}, \bibinfo {author} {\bibfnamefont {X.}~\bibnamefont {Lu}}, \bibinfo
  {author} {\bibfnamefont {G.~K.~W.}\ \bibnamefont {Koon}}, \bibinfo {author}
  {\bibfnamefont {A.~H.~C.}\ \bibnamefont {Neto}}, \bibinfo {author}
  {\bibfnamefont {B.}~\bibnamefont {\"{O}zyilmaz}}, \bibinfo {author}
  {\bibfnamefont {Q.}~\bibnamefont {Xiong}}, \ and\ \bibinfo {author}
  {\bibfnamefont {S.~Y.}\ \bibnamefont {Quek}},\ }\href@noop {} {\bibfield
  {journal} {\bibinfo  {journal} {Nano Letters}\ }\textbf {\bibinfo {volume}
  {15}},\ \bibinfo {pages} {3931} (\bibinfo {year} {2015})}\BibitemShut
  {NoStop}%
\bibitem [{\citenamefont {Choi}\ and\ \citenamefont
  {Strano}(2007)}]{choi2007solvatochromism}%
  \BibitemOpen
  \bibfield  {author} {\bibinfo {author} {\bibfnamefont {J.~H.}\ \bibnamefont
  {Choi}}\ and\ \bibinfo {author} {\bibfnamefont {M.~S.}\ \bibnamefont
  {Strano}},\ }\href@noop {} {\bibfield  {journal} {\bibinfo  {journal}
  {Applied Physics Letters}\ }\textbf {\bibinfo {volume} {90}},\ \bibinfo
  {pages} {3114} (\bibinfo {year} {2007})}\BibitemShut {NoStop}%
\bibitem [{\citenamefont {Lin}\ \emph {et~al.}(2014)\citenamefont {Lin},
  \citenamefont {Ling}, \citenamefont {Yu}, \citenamefont {Huang},
  \citenamefont {Hsu}, \citenamefont {Lee}, \citenamefont {Kong}, \citenamefont
  {Dresselhaus},\ and\ \citenamefont {Palacios}}]{Lin14}%
  \BibitemOpen
  \bibfield  {author} {\bibinfo {author} {\bibfnamefont {Y.}~\bibnamefont
  {Lin}}, \bibinfo {author} {\bibfnamefont {X.}~\bibnamefont {Ling}}, \bibinfo
  {author} {\bibfnamefont {L.}~\bibnamefont {Yu}}, \bibinfo {author}
  {\bibfnamefont {S.}~\bibnamefont {Huang}}, \bibinfo {author} {\bibfnamefont
  {A.~L.}\ \bibnamefont {Hsu}}, \bibinfo {author} {\bibfnamefont {Y.-H.}\
  \bibnamefont {Lee}}, \bibinfo {author} {\bibfnamefont {J.}~\bibnamefont
  {Kong}}, \bibinfo {author} {\bibfnamefont {M.~S.}\ \bibnamefont
  {Dresselhaus}}, \ and\ \bibinfo {author} {\bibfnamefont {T.}~\bibnamefont
  {Palacios}},\ }\href@noop {} {\bibfield  {journal} {\bibinfo  {journal} {Nano
  Letters}\ }\textbf {\bibinfo {volume} {14}},\ \bibinfo {pages} {5569}
  (\bibinfo {year} {2014})}\BibitemShut {NoStop}%
\bibitem [{\citenamefont {Li}\ and\ \citenamefont
  {Appelbaum}(2014)}]{li2014electrons}%
  \BibitemOpen
  \bibfield  {author} {\bibinfo {author} {\bibfnamefont {P.}~\bibnamefont
  {Li}}\ and\ \bibinfo {author} {\bibfnamefont {I.}~\bibnamefont {Appelbaum}},\
  }\href@noop {} {\bibfield  {journal} {\bibinfo  {journal} {Physical Review
  B}\ }\textbf {\bibinfo {volume} {90}},\ \bibinfo {pages} {115439} (\bibinfo
  {year} {2014})}\BibitemShut {NoStop}%
\bibitem [{\citenamefont {Baba}\ \emph {et~al.}(1991)\citenamefont {Baba},
  \citenamefont {Nakamura}, \citenamefont {Shibata},\ and\ \citenamefont
  {Morita}}]{baba1991photoconduction}%
  \BibitemOpen
  \bibfield  {author} {\bibinfo {author} {\bibfnamefont {M.}~\bibnamefont
  {Baba}}, \bibinfo {author} {\bibfnamefont {Y.}~\bibnamefont {Nakamura}},
  \bibinfo {author} {\bibfnamefont {K.}~\bibnamefont {Shibata}}, \ and\
  \bibinfo {author} {\bibfnamefont {A.}~\bibnamefont {Morita}},\ }\href@noop {}
  {\bibfield  {journal} {\bibinfo  {journal} {Japanese Journal of Applied
  Physics}\ }\textbf {\bibinfo {volume} {30}},\ \bibinfo {pages} {L1178}
  (\bibinfo {year} {1991})}\BibitemShut {NoStop}%
\bibitem [{\citenamefont {Cardona}\ and\ \citenamefont
  {Thewalt}(2005)}]{cardona2005isotope}%
  \BibitemOpen
  \bibfield  {author} {\bibinfo {author} {\bibfnamefont {M.}~\bibnamefont
  {Cardona}}\ and\ \bibinfo {author} {\bibfnamefont {M.~L.~W.}\ \bibnamefont
  {Thewalt}},\ }\href@noop {} {\bibfield  {journal} {\bibinfo  {journal}
  {Reviews of Modern Physics}\ }\textbf {\bibinfo {volume} {77}},\ \bibinfo
  {pages} {1173} (\bibinfo {year} {2005})}\BibitemShut {NoStop}%
\bibitem [{\citenamefont {Vi{\~n}a}\ \emph {et~al.}(1984)\citenamefont
  {Vi{\~n}a}, \citenamefont {Logothetidis},\ and\ \citenamefont
  {Cardona}}]{vina1984temperature}%
  \BibitemOpen
  \bibfield  {author} {\bibinfo {author} {\bibfnamefont {L.}~\bibnamefont
  {Vi{\~n}a}}, \bibinfo {author} {\bibfnamefont {S.}~\bibnamefont
  {Logothetidis}}, \ and\ \bibinfo {author} {\bibfnamefont {M.}~\bibnamefont
  {Cardona}},\ }\href@noop {} {\bibfield  {journal} {\bibinfo  {journal}
  {Physical Review B}\ }\textbf {\bibinfo {volume} {30}},\ \bibinfo {pages}
  {1979} (\bibinfo {year} {1984})}\BibitemShut {NoStop}%
\bibitem [{\citenamefont {Cardona}\ and\ \citenamefont
  {Kremer}(2014)}]{cardona2014temperature}%
  \BibitemOpen
  \bibfield  {author} {\bibinfo {author} {\bibfnamefont {M.}~\bibnamefont
  {Cardona}}\ and\ \bibinfo {author} {\bibfnamefont {R.~K.}\ \bibnamefont
  {Kremer}},\ }\href@noop {} {\bibfield  {journal} {\bibinfo  {journal} {Thin
  Solid Films}\ }\textbf {\bibinfo {volume} {571}},\ \bibinfo {pages} {680}
  (\bibinfo {year} {2014})}\BibitemShut {NoStop}%
\bibitem [{\citenamefont {Dey}\ \emph {et~al.}(2013)\citenamefont {Dey},
  \citenamefont {Paul}, \citenamefont {Bylsma}, \citenamefont {Karaiskaj},
  \citenamefont {Luther}, \citenamefont {Beard},\ and\ \citenamefont
  {Romero}}]{dey2013origin}%
  \BibitemOpen
  \bibfield  {author} {\bibinfo {author} {\bibfnamefont {P.}~\bibnamefont
  {Dey}}, \bibinfo {author} {\bibfnamefont {J.}~\bibnamefont {Paul}}, \bibinfo
  {author} {\bibfnamefont {J.}~\bibnamefont {Bylsma}}, \bibinfo {author}
  {\bibfnamefont {D.}~\bibnamefont {Karaiskaj}}, \bibinfo {author}
  {\bibfnamefont {J.~M.}\ \bibnamefont {Luther}}, \bibinfo {author}
  {\bibfnamefont {M.~C.}\ \bibnamefont {Beard}}, \ and\ \bibinfo {author}
  {\bibfnamefont {A.~H.}\ \bibnamefont {Romero}},\ }\href@noop {} {\bibfield
  {journal} {\bibinfo  {journal} {Solid State Communications}\ }\textbf
  {\bibinfo {volume} {165}},\ \bibinfo {pages} {49} (\bibinfo {year}
  {2013})}\BibitemShut {NoStop}%
\bibitem [{\citenamefont {Lian}\ \emph {et~al.}(2006)\citenamefont {Lian},
  \citenamefont {Yang}, \citenamefont {Thewalt}, \citenamefont {Lauck},\ and\
  \citenamefont {Cardona}}]{lian2006effects}%
  \BibitemOpen
  \bibfield  {author} {\bibinfo {author} {\bibfnamefont {H.~J.}\ \bibnamefont
  {Lian}}, \bibinfo {author} {\bibfnamefont {A.}~\bibnamefont {Yang}}, \bibinfo
  {author} {\bibfnamefont {M.~L.~W.}\ \bibnamefont {Thewalt}}, \bibinfo
  {author} {\bibfnamefont {R.}~\bibnamefont {Lauck}}, \ and\ \bibinfo {author}
  {\bibfnamefont {M.}~\bibnamefont {Cardona}},\ }\href@noop {} {\bibfield
  {journal} {\bibinfo  {journal} {Physical Review B}\ }\textbf {\bibinfo
  {volume} {73}},\ \bibinfo {pages} {233202} (\bibinfo {year}
  {2006})}\BibitemShut {NoStop}%
\bibitem [{\citenamefont {Aierken}\ \emph {et~al.}(2015)\citenamefont
  {Aierken}, \citenamefont {{\c{C}}ak{\i}r}, \citenamefont {Sevik},\ and\
  \citenamefont {Peeters}}]{aierken2015thermal}%
  \BibitemOpen
  \bibfield  {author} {\bibinfo {author} {\bibfnamefont {Y.}~\bibnamefont
  {Aierken}}, \bibinfo {author} {\bibfnamefont {D.}~\bibnamefont
  {{\c{C}}ak{\i}r}}, \bibinfo {author} {\bibfnamefont {C.}~\bibnamefont
  {Sevik}}, \ and\ \bibinfo {author} {\bibfnamefont {F.~M.}\ \bibnamefont
  {Peeters}},\ }\href@noop {} {\bibfield  {journal} {\bibinfo  {journal}
  {Physical Review B}\ }\textbf {\bibinfo {volume} {92}},\ \bibinfo {pages}
  {081408} (\bibinfo {year} {2015})}\BibitemShut {NoStop}%
\bibitem [{\citenamefont {Leroux}\ \emph {et~al.}(1999)\citenamefont {Leroux},
  \citenamefont {Grandjean}, \citenamefont {Beaumont}, \citenamefont {Nataf},
  \citenamefont {Semond}, \citenamefont {Massies},\ and\ \citenamefont
  {Gibart}}]{leroux1999temperature}%
  \BibitemOpen
  \bibfield  {author} {\bibinfo {author} {\bibfnamefont {M.}~\bibnamefont
  {Leroux}}, \bibinfo {author} {\bibfnamefont {N.}~\bibnamefont {Grandjean}},
  \bibinfo {author} {\bibfnamefont {B.}~\bibnamefont {Beaumont}}, \bibinfo
  {author} {\bibfnamefont {G.}~\bibnamefont {Nataf}}, \bibinfo {author}
  {\bibfnamefont {F.}~\bibnamefont {Semond}}, \bibinfo {author} {\bibfnamefont
  {J.}~\bibnamefont {Massies}}, \ and\ \bibinfo {author} {\bibfnamefont
  {P.}~\bibnamefont {Gibart}},\ }\href@noop {} {\bibfield  {journal} {\bibinfo
  {journal} {Journal of Applied Physics}\ }\textbf {\bibinfo {volume} {86}},\
  \bibinfo {pages} {3721} (\bibinfo {year} {1999})}\BibitemShut {NoStop}%
\bibitem [{\citenamefont {Fang}\ \emph {et~al.}(2015)\citenamefont {Fang},
  \citenamefont {Wang}, \citenamefont {Sun}, \citenamefont {Lu}, \citenamefont
  {Deng}, \citenamefont {Ma}, \citenamefont {Jiang}, \citenamefont {Jia},
  \citenamefont {Wang}, \citenamefont {Zhou},\ and\ \citenamefont
  {Chen}}]{fang2015investigation}%
  \BibitemOpen
  \bibfield  {author} {\bibinfo {author} {\bibfnamefont {Y.}~\bibnamefont
  {Fang}}, \bibinfo {author} {\bibfnamefont {L.}~\bibnamefont {Wang}}, \bibinfo
  {author} {\bibfnamefont {Q.}~\bibnamefont {Sun}}, \bibinfo {author}
  {\bibfnamefont {T.}~\bibnamefont {Lu}}, \bibinfo {author} {\bibfnamefont
  {Z.}~\bibnamefont {Deng}}, \bibinfo {author} {\bibfnamefont {Z.}~\bibnamefont
  {Ma}}, \bibinfo {author} {\bibfnamefont {Y.}~\bibnamefont {Jiang}}, \bibinfo
  {author} {\bibfnamefont {H.}~\bibnamefont {Jia}}, \bibinfo {author}
  {\bibfnamefont {W.}~\bibnamefont {Wang}}, \bibinfo {author} {\bibfnamefont
  {J.}~\bibnamefont {Zhou}}, \ and\ \bibinfo {author} {\bibfnamefont
  {H.}~\bibnamefont {Chen}},\ }\href@noop {} {\bibfield  {journal} {\bibinfo
  {journal} {Scientific Reports}\ }\textbf {\bibinfo {volume} {5}},\ \bibinfo
  {pages} {12718} (\bibinfo {year} {2015})}\BibitemShut {NoStop}%
\end{thebibliography}%


\end{document}